\documentclass[11pt]{article}
\begin{document}

\title{
Symmetries of Integro-Differential Equations: A Survey of Methods
Illustrated by  the Benney Equations}
\author{N.~H.~IBRAGIMOV \\
 {\footnotesize\textit{ International Research Centre ALGA: Advances
 in Lie Group Analysis,}} \\
 {\footnotesize\textit{ Department of Mathematics,
 Blekinge Institute of Technology,}} \\
 {\footnotesize\textit{ 371 79 Karlskrona, Sweden}} \\
\and {V.~F.~KOVALEV} \\
 {\footnotesize\textit{ Institute for Mathematical Modelling, Russian Academy
 of Sciences,}} \\
 {\footnotesize\textit{ Miusskaya Sq., 4A, Moscow, 125047, Russia}} \\
 \and {V.~V.~PUSTOVALOV} \\
 {\footnotesize\textit{P.~N.~Lebedev Physical Institute,  Russian Academy of
 Sciences,}} \\ {\footnotesize\textit{Leninski Pr., 53, Moscow,
 117333, Russia}} }
  \date{}
  \maketitle

 \begin{abstract}
    Classical Lie group theory provides a universal tool for
    calculating symmetry groups for systems of differential equations.
    However Lie's method is not as much effective in the case of
    integral or integro-differential equations as well as in the case
    of infinite systems of differential equations.
    \par
    This paper is aimed to survey the modern approaches to symmetries
    of integro-differential equations. As an illustration, an infinite
    symmetry Lie algebra is calculated for a system of
    integro-differential equations, namely the well-known Benney
    equations. The crucial idea is to look for symmetry generators in
    the form of canonical Lie-B\"acklund operators.
 \end{abstract}

 \section{Introduction}

The major obstacle for the application of Lie's infinitesimal
techniques to integro-differential equations or infinite systems
of differential equations is that the {\textit{frames}} (see, e.g.
\cite{Ibr99} or \cite{Ibr00}) of these equations are not locally
defined in the space of differential functions. In consequence,
the crucial idea of {\textit splitting} of determining equations
into over-determined systems, commonly used in the classical Lie
group analysis, fails.

\subsection{Different forms of the Benney equations}

The Benney equations referred to by the name of the author of a
pioneering work \cite{Ben73} appear in long wavelength
hydrodynamics of an ideal incompressible fluid of a finite depth
in a gravitational field. From the group theoretical point of view
they are of particular interest due to the existence of an
infinite set of conservation laws obtained in \cite{Ben73}. The
latter property of the Benney equations emphasizes their
significance that goes far beyond an interesting example of an
integrable system of hydrodynamic equations.
\par
In practice, the Benney equations are used in various
representation. One of them is the kinetic Benney equation (a
kinetic equation with a self-consistent field):
 \begin{equation}
 f_t + vf_x -A_{x}^{0}f_v =0, \qquad
 A^0(t,x)=\int\limits_{-\infty}^{+\infty} f(t,x,v) dv.
 \label{ben1}
 \end{equation}
This equation appears as a unique representative of a set of
hierarchy of kinetic equations of Vlasov-type \cite{Kras89}. A
detailed study of its group properties will lead to better
understanding of the symmetry properties of kinetic equations of
collisionless plasma, viz. the Vlasov-Maxwell equations (see also
Appendix 1) that have both theoretical and practical interest,
e.g. while dealing with controlled nuclear fusion programme.
\par
Another form of the Benney equations is an infinite set of coupled
equations
 \begin{equation}
 A_{t}^{i} + A_{x}^{i+1} + i A_{x}^{0}A^{i-1} = 0, \qquad i \geq 0
 \label{ben2}
 \end{equation}
for a countable set of functions $A^i$ of two independent
variables, time $t$ and the spatial coordinate $x$. In terms of
hydrodynamics these functions appear as averaged values (with
respect to the depth) of integer powers $ i\geq0 $ of the
horizontal component of the liquid flow velocity. The
corresponding integrals that describe this averaging are taken
over the vertical coordinate in the limits from the flat bottom up
to the free liquid surface. Solutions, Hamiltonian structure and
conservation laws for the equations (\ref{ben2}) were discussed in
details in \cite{Kup77,Kup78}.
\par
>From the kinetic point of view the system (\ref{ben2}) can be
treated as a system of equations for moments of the distribution
function $f$ that obeys the kinetic Benney equation (\ref{ben1})
 \begin{equation}
 A^i(t,x)=\int\limits_{-\infty}^{+\infty}  v^i f d v, \qquad
 i\geq 0 \,.
 \label{moment}
 \end{equation}
This fact with the explicit formulation of the Benney equation
(\ref{ben1}) was first stated independently in \cite{Zak80,Gib81}.
The Lagrangian change of the Euler velocity $v,$
 \begin{equation}
 v=V(t,x,u)
 \label{lagr-v}
 \end{equation}
yields one more representation for Benney equations (\ref{ben1}):
 \begin{equation}
 f_t + Vf_x = 0\,, \quad V_t + VV_x = -A^{0}_{x} \,, \quad
 A^0(t,x)=\int V_u f(t,x,u) d u .
 \label{ben3}
 \end{equation}
The equations (\ref{ben3}) are readily converted into the
hydrodynamic-type form
 \begin{equation}
 n_t + (n V)_x = 0\,, \quad V_t + VV_x = -A^{0}_{x} \,, \quad
 A^0=\int n(t,x,u) d u ,
 \label{ben4}
 \end{equation}
if one employs the ``density" $n$ depending on the Lagrangian
velocity $u$:
 \begin{equation}
 n = f(t,x,u) V_u \,.
 \label{lagr-n}
 \end{equation}
Using the form (\ref{ben4}) of the Benney equations an infinite
set of conservation laws were constructed in \cite{Zak80} with the
densities regarded as functions of the Lagrangian velocity~$u$.
\par
Furthermore, we will rewrite the integro-differential Benney
equations (\ref{ben1}) in the form of differential equations by
introducing the following nonlocal variables:
 \begin{equation}
 g = \int\limits_{-\infty}^{v} f d v, \quad
 h = \int\limits^{+\infty}_{v} f d v.
 \label{nloc1}
 \end{equation}
In terms of the latter variables the equations (\ref{ben1}) are
written as
 \begin{equation}
 f_t + v f_x - ( g_x + h_x ) f_v =0, \quad
 g_v = f \,, \quad h_v = - f \,.
 \label{ben5}
 \end{equation}
\par
The knowledge of the complete Lie-B\"acklund symmetry for the
Benney equations in different representations (\ref{ben1}),
(\ref{ben2}), (\ref{ben3}), (\ref{ben4}) and (\ref{ben5}) can
clarify the question of structure of solutions and conservation
laws for these equations. This statement is partially confirmed by
the fact that one of the main results of the works
\cite{Kup77,Kup78}, namely the higher order Benney equations, can
be re-formulated in terms of the first order Lie-B\"acklund group,
admitted by the system (\ref{ben2}). Unfortunately, the complete
description of the Lie-B\"acklund symmetry for the equations
(\ref{ben2}) is not available in the literature. The goal of this
paper is to contribute to this problem by calculating an infinite
(countable) part of the Lie point symmetries of the moment
equations (\ref{ben2}).

\section{Generalities}

Here, we briefly discuss different known approaches to calculating
symmetry groups for integro-differential equations. Loosely
speaking, these approaches can be divided into two large groups:
indirect and direct methods.

Algorithms of the first group rest on the possibility to replace
in any way input nonlocal (integro-differential) equations by a
system of differential equations. Then the resulting system of
differential equations is analyzed using standard methods of a
classical Lie group analysis. Here we point on two different ways
of reducing nonlocal equations to differential ones.

\par
\subsection{Indirect methods}

A. {\textit{Method of moments}}\\[1ex]
 In this approach, the system of basic \textit{integro-differential}
equations for a function $f$ (e.g., the kinetic equation
(\ref{ben1})), that usually contains nonlocal terms depending on
moments (\ref{moment}) of this function, is reduced to an infinite
system of \textit{differential} equations for these moments (in
our case this is the system (\ref{ben2})). The admitted symmetry
group is then calculated using the traditional methods of Lie
group analysis for any \textit{finite} subsystem of $k$ equations
of this general \textit{infinite} system. Then an intersection of
all admitted groups is defined and a transition to the limiting
case $k \to \infty$ is fulfilled. The resultant algebra of group
generators thus obtained is used to reconstruct the algebra with
the original function $f$ directly involved. This procedure
usually employs the explicit form of finite group transformations
for moments (\ref{moment}) and the relations between $A^i$ and the
function $f$. The last step is not trivial in any case as there
may exist different representation for the equations for the
moments $A^i$ and the resultant group depends on the form of this
representation. Hence, the transition from the algebra of group
generators for the moments representation back to the symmetry of
original equations formulated in terms of input functions form a
special problem which we are going to discuss in the section
\ref{prolong}. The above described algorithm was realized to
calculate Lie point symmetry group for Vlasov-Maxwell equations in
plasma theory \cite{Taranov-74}--\cite{Taranov-79} and for Benney,
Vlasov-type and Boltzmann-type kinetic equations
\cite{Bun-mjg-82,Bun-vestmgu-82}, \cite{Kras89}.\\[2ex]
 B. {\textit{ Method of boundary-differential equations}}\\[1ex]
 The following method was developed in \cite{Chetv-AAM95,Chetv-AMST95} on
basis of the concept of covering and applied to a coagulation
kinetic equation. In this case each definite integral in the
equation is replaced by the corresponding difference of values of
the antiderivative on a boundary sets. The integro-differential
equation takes the boundary (or functional) differential form. As
shown in  \cite{Chetv-AAM95,Chetv-AMST95} a geometric theory of
boundary differential equations can be constructed in just the
same way as the analogous theory of differential equations thus
allowing to define and compute not only classical but higher
symmetries as well. It is noted that the elimination of integrals
by introducing potentials depends on the choice of potential
variables and can be executed in different ways. Hence the result
of group calculation and its dimension is influenced by the form
of potential variables involved.

\subsection{Direct methods}

Direct methods of finding symmetries were  developed in
\cite{Mel-Gr-86}-\cite{Mel-prl-98} (see also \cite{bi}) and
\cite{KKP92,KKP93} and applied to find symmetries of kinetic
Boltzmann equation, the equations of motion of viscoelastic medium
and Vlasov-Maxwell equations of plasma theory. To extend the
classical Lie algorithm to integro-differential equations it
appears necessary to resolve several problems. First, one should
define the local one-parameter transformation group $G$ for the
nonlocal (integro-differential) equations and formulate the
invariance criteria that lead to determining equations.

Let an integro-differential equation under consideration be
expressed as a zero equality for some functional (here we indicate
only one argument for a function $f$)
 \begin{equation}
 F\left[ f(x) \right]=0\,, \label{gen}
 \end{equation}
and let $G$ be a local one-parameter group that transforms $f$ to
$\tilde f(x)$,
 \begin{equation}
 \tilde f(x)= f + a \mbox{\ae} + o(a^2)\,, \quad \tilde x = x\,,
 \label{group}
 \end{equation}
Here we use the canonical group representation hence independent
variables $x$ do not vary.  Then the local group $G$ of point
transformations (\ref{group}) is called a symmetry group of
integro-differential equations (\ref{gen}) iff for any $a$ the
function $F$ does not vary \cite{Mel-Gr-87},
 \begin{equation}
 F\left[\tilde f(x) \right]=0\,. \label{inv}
 \end{equation}
Differentiating (\ref{inv}) with respect to group parameter $a$
and assuming $a \to 0$ gives the determining equations. In
contrast to the case of input differential equations these
determining equations are in general also integro-differential.

The invariance criterion for $F$ with respect to the admitted
group can be expressed in an infinitesimal form using the
canonical group operator $Y$,
 \begin{equation}
 {\left. Y F \rule[-7pt]{0pt}{19pt}\right|}_{F=0}=0\,, \quad
 \mbox{where} \quad Y\equiv \int\mbox{d}y
 {\mbox{\ae}}\left(y\right)\, \frac{{\delta}}
 {{\delta}f\left(y\right)}\,.
 \label{inv-cri}\end{equation}
Here with the goal to generalize the action of a canonical group
operator not only on differential functions but on
\textit{functionals} as well we use variational differentiation in
the definition of $Y$ \cite{KKP93,KKP98}. One can verify by direct
calculation that the action of $Y$ on any differential function
and its derivatives, e.g., $f$ and $f_x, \, \ldots $  produces the
usual result: $Y f = \mbox{\ae}$, $ Y f_x = D_x(\mbox{\ae})$ and
so on. Hence, if $F$ describe  usual differential equations then
formulas (\ref{inv-cri}) lead to standard local determining
equations, while for $F$ having the form of integro-differential
equations formulas (\ref{inv-cri}) can be treated as
\textit{nonlocal} determining equations as they depend both on
local and nonlocal variables.

In order to find solutions of determining equations one can use
different approaches, e.g. expanding coordinates of group
generator into formal power series and equating coefficients of
various powers \cite{Mel-Gr-86,Mel-Gr-87}. However there exists a
more traditional way. As we treat local and nonlocal variables in
determining equations as independent it is possible to separate
these equations into local and nonlocal. The procedure of solving
local determining equations is fulfilled in a standard way using
Lie algorithm based on splitting the system of over-determined
equations with respect to local variables and their derivatives.
As a result we get expressions for coordinates of group generator
that define the so-called group of \textit{intermediate} symmetry
\cite{KKP93}. In the similar manner the solution of nonlocal
determining equations is fulfilled using the information borrowed
from an intermediate symmetry and by ``variational" splitting of
nonlocal determining equations using the procedure of variational
differentiation. Therefore, the algorithm of finding symmetries of
nonlocal equations appears as an algorithmic procedure that
consists of a sequence of several steps: a) defining the set of
local group variables, b) constructing determining equations on
basis of the infinitesimal criterion of invariance, that employs
the generalization of the definition of the canonical operator, c)
separating determining equations into local and nonlocal, d)
solving local determining equations using a standard Lie
algorithm, e) solving nonlocal determining equations using the
procedure of variational differentiation.

In the next sections,  the above methods are applied to the Benney
equations.

\section{Lie subgroup and Lie-B\"acklund group: statement of the problem}
\subsection{Lie subgroup: direct method of calculation}
A Lie subgroup, admitted by the kinetic Benney equation
(\ref{ben1}) in the space of four group variables
 \begin{equation} t,\  x, \ v, \  f
 \label{var1} \end{equation}
is defined by five basic infinitesimal operators
 \begin{equation}
 \begin{array}{l}
 \displaystyle{
 X_1 = \partial_{t}, \quad
 X_2 = \partial_{x}, \quad
 X_3 = t \partial_{x} + \partial_{v},} \\
 \mbox{} \\
 \displaystyle{
 X_4 = t \partial_{t} - v\partial_{v} - f\partial_{f}, \quad
 X_5 = x \partial_{x} + v\partial_{v} + f\partial_{f}\,.}
 \end{array}
 \label{ben-gen}
 \end{equation}
With the less computation difficulties this group can be obtained
using the approach developed in \cite{KKP92,KKP93,KKP98} in
application to group analysis of Vlasov-Maxwell equations in
plasma theory (see also Chapter 16 in Handbook \cite{CRC}).
\par
Here we demonstrate the application of the general scheme to
Benney equations (\ref{ben1}). The merit of this direct approach
is the possibility to present the criterion of invariance with
respect to local one-parameter point group transformations in a
standard infinitesimal form. For the Benney equations the point
symmetry group generator has the form
\begin{equation}
\label{gen-vl}
 X= \xi^{1} \partial_{t} +
    \xi^{2} \partial_{x} +
    \xi^{3} \partial_{v} +
    \eta^{1} \partial_{f} +
    \eta^{2} \partial_{A^{0}}  \,,
\end{equation}
where the coordinates $\xi$ and $\eta$ depend on $t$, $x$, $v$,
$f$ and $A^0$.

In the canonical form this operator is written:
 \begin{equation}
 Y = {\mbox{\ae}}^{1} {\partial}_{f}
   + {\mbox{\ae}}^{2}{\partial}_{A^0} \,,
  \label{can-oper}
  \end{equation}
where
 \[
 {\mbox{\ae}}^{1} = {\eta}^{1} - {\xi}^{1}f_t - {\xi}^{2}f_x
                  - {\xi}^{3}f_v \, , \qquad
 {\mbox{\ae}}^{2} = {\eta}^{2} - {\xi}^{1}A^0_t - {\xi}^{2}A^0_x \,,
 \]
and its action on any function or functional should be understood
in generalized sense as in (\ref{inv-cri}).  Applying the
canonical group operator to the joint system of basic equations
(\ref{ben1}) and one more ``evident" equation, that expresses the
fact that the moment of $f$ does not depend upon $v$
\begin{equation} A^0_v=0 \,,  \label{der_v} \end{equation}
gives the system of determining equations
 \begin{equation}
 {D}_{t}\left({\mbox{\ae}}^{1}\right) +
 v{D}_{x}\left({\mbox{\ae}}^{1}\right) -
 A^0_x{D}_{v}\left({\mbox{\ae}}^{1}\right) -
 {D}_{x}\left({\mbox{\ae}}^{2}\right){f}_{v} = 0\,, \quad
 {D}_{v}\left({\mbox{\ae}}^{2}\right)=0\, ;
 \label{deteq1}
 \end{equation}
 \begin{equation}
 {\mbox{\ae}}^{2}=\int {\mbox{\ae}}^{1} d v ,
 \label{deteq2}
 \end{equation}
which should be solved in view of the the complete set of basic
equations (\ref{ben1}), (\ref{der_v}).
\par
In solving determining equations (\ref{deteq1}) group variables
 \begin{equation}
 \{ f, \, A^0, \,f_x, \, A^0_x, \,f_v, \,\ldots \}
 \label{var2}
 \end{equation}
are treated as independent ones. This assumption separates
determining equations into local equations (\ref{deteq1}) and
nonlocal equation (\ref{deteq2}). Local determining equations are
solved in a standard way using the computational algorithm of Lie
group analysis. Then the functions $\xi$ and $\eta$ thus obtained
define the so-called \textit{intermediate} symmetry~\cite{KKP93}
 \begin{equation}
 \begin{array}{l}
 \displaystyle{
 {\xi}^1  =  {\xi}^1(t)\,, \quad
 {\xi}^2  =  \frac{x}{2}\, {\xi}^1_t (t)
             + \alpha x + \beta (t)\,, \quad
 {\xi}^3  =   \alpha v - \frac{v}{2}\, {\xi}^1_t
              +\frac{x}{2}\, {\xi}^1_{tt} + {\beta}_{t}\,,} \\
 \mbox{}\\
 \displaystyle{
 {\eta}^1  = {\eta}^1 (f) \,, \quad
 {\eta}^2  = \gamma (t) + (2 \alpha - \xi^1_{t}) A^0 -
              x {\beta}_{tt} - \frac{x^2}{4}\, {\xi}^1_{ttt}\,.}
 \end{array}
 \label{coords}
 \end{equation}
Here ${\xi}^1(t)$, $\beta(t)$, $\gamma(t)$ and ${\eta}^1(f)$ are
arbitrary functions of their arguments and $\alpha$ and $\nu$ are
constants.

Now let us turn to the solution of the nonlocal determining
equation (\ref{deteq2}) that we rewrite in the following form
 \begin{equation}
 {\eta}^{2} - {\xi}^{1} A^0_t - {\xi}^{2} A^0_x  = \int \left(
  \, {\eta}^{1} - {\xi}^{1} f_t - {\xi}^{2} f_x - {\xi}^{3}f_v
  \right) d v \,.
 \label{nl-det2}
 \end{equation}
As in the case of the local determining equations (\ref{deteq1}),
the latter should be solved in view of the original equations
(\ref{ben1}), (\ref{der_v}). Hence, calculating the derivatives
$A^0_t$, $A^0_x$ and $f_t$ from the basic equation (\ref{ben1})
and inserting them into (\ref{nl-det2}) and in view of the above
expressions for coordinates $\xi$ and $\eta$ we obtain the
following nonlocal determining equation
 \begin{equation}
 \int\limits_{-\infty}^{+\infty} \left[\ {\eta}^1(f) - \left( \alpha -
 \frac{1}{2}\, {\xi}^1_{t} \right) f \right] d v =
   \gamma - x {\beta}_{tt} -\frac{x^2}{4} \, {\xi}^1_{ttt} \,.
 \label{nl-det3}
 \end{equation}
As any determining equation, (\ref{nl-det3}) is the equality with
respect to all group variables that appear in this equation.
Therefore differentiating it with respect to any group variables
also leads to equalities. Hence, nonlocal determining equation can
be split with respect to independent group variable $f$ using the
variational differentiation.
Since the right-hand part of (\ref{nl-det3}) does not depend on
$f$, it reduces after differentiation and the remaining terms are
written as
 \begin{equation}
 \frac{\delta}{\delta f(v')}
  \int\limits_{-\infty}^{+\infty} \left[\ {\eta}^1(f) - \left( \alpha -
 \frac{1}{2}\, {\xi}^1_{t} \right) f \right] d v = 0 .
 \label{nl-det4}
 \end{equation}
It is essential that the nonlocal determining equation
(\ref{nl-det2}) should be solved simultaneously with its
differential consequence, i.e. any solution of (\ref{nl-det4})
must appear as the solution of (\ref{nl-det3}). Introducing the
variational derivative $(\delta / \delta f(v'))$ inside the
integral over $v$
 \[
   \int\limits_{-\infty}^{+\infty}
   \left[\ {\eta}^1_f - \left( \alpha -
 \frac{1}{2}\, {\xi}^1_{t} \right) \right]
    \frac{\delta f(v)}{\delta f(v')} \, d v = 0 ,
 \]
and eliminating integration over $v$ with the help of the Dirac
delta-function,
 \[
 \frac{\delta f(v)}{\delta f(v')} = \delta (v - v') \,,
 \]
one comes to the first order differential equation for $\eta^1$
 \[ {\eta}^1_f - \left( \alpha -
 \frac{1}{2}\, {\xi}^1_{t} \right)  = 0 \,, \]
 which gives the linear dependence of
$\eta^1$ upon $f$,
 \begin{equation}
 \eta^1 = f \left( \alpha -  \frac{1}{2}\, \xi^1_t \right) + C\,,
 \label{coord}
 \end{equation}
with some constant $C$. Substituting this result back into
(\ref{nl-det3}) yields the zero-value of this constant provided
integral has finite value. As $\eta^1$ does not depend on $t$ then
differentiating (\ref{coord}) with respect to $t$ gives
$\xi^{1}_{tt}=0$. Differentiating (\ref{nl-det3}) with respect to
$x$ gives two more equations,
 $$\gamma =\beta_{tt} = 0\,. $$
Solving these equations gives final expressions for coordinates
$\xi$ and $\eta$:
 \begin{equation}
 \begin{array}{l}
 \displaystyle{
 \xi^{1} = c^1 +c^4 t \,, \quad
 \xi^{2} = c^2 + c^3 t + c^5 x \,, \quad
 \xi^{3} = c^3 + (- c^4 +c^5) v \,, }\\
 \mbox{}\\
  \displaystyle{
 \eta^{1} = (- c^4 +c^5) f \,, \quad
 \eta^{2} = 2(- c^4 +c^5) A^0 \,.}
 \end{array}
 \label{coord3}
 \end{equation}
These coordinates give rise to the five-dimensional Lie algebra
with generators given by (\ref{ben-gen}).  One can see that terms
proportional to $A^0$ which come from $\mbox{\ae}^2$ are omitted
in $X_4$, $X_5$ as they appear as the result of
\textit{prolongation} \cite{KKP96} of these operators \textit{on a
nonlocal variable} $A^0$. This procedure for any of operators
$X_4$, $X_5$ is described  in a concise form in the next section.

\subsection{Extension of Lie point symmetry generators to
             nonlocal variables \label{prolong}}

To fulfill the procedure of prolongation of the Lie point symmetry
generators one should first rewrite the operator, say $X_5$, in a
canonical form
 \begin{equation}
 Y_{5} = {\mbox{\ae}}^1 \partial_f \,, \quad
         {\mbox{\ae}}^1= ( - x f_x - v f_v + f)\,.
 \label{canonic2}
 \end{equation}
Then formally prolong this operator on the nonlocal variable $A^0$
 \begin{equation}
 Y_{5} + \mbox{\ae}^2 \equiv {\mbox{\ae}}^1 \partial_f
                      + {\mbox{\ae}}^2 \partial_{A^0} \,.
 \label{prolong1}
 \end{equation}
The integral relation between ${\mbox{\ae}}^1$ and
${\mbox{\ae}}^2$ is obtained by applying the generator
(\ref{prolong1}) to the second equation in (\ref{ben1}) that is
treated here as the \textit{definition} of $A^0$. This relation
here coincides with nonlocal determining equation (\ref{deteq2}).
Substituting ${\mbox{\ae}}^1$ from (\ref{canonic2}) in
(\ref{deteq2}) and calculating integrals obtained (integrating by
paths) gives the desired coordinate ${\mbox{\ae}}^2$
 \begin{equation}
 \mbox{\ae}^2 = ( 2 A^0 - x A^0_x )\,.
 \label{canonic3}
 \end{equation}
Inserting this into (\ref{prolong1}) and returning back to the
non-canonical representation we get the following generator
 \begin{equation}
 X_{5} = x\partial_x + v \partial_v
       + f \partial_f + 2 A^0 \partial_{A^0} \,,
 \label{oper2}
 \end{equation}
that correlates with the result (\ref{coord3}).
\par
Prolongation of infinitesimal operators (\ref{ben-gen}) on
nonlocal variables (\ref{moment}) extends the set of group
variables (\ref{var1}) up to a countable set
 \begin{equation} t, \ x, \ v, \ f, \ A^0, \ \ldots , \ A^i, \ \ldots \,.
 \label{var3}
 \end{equation}
In the latter case infinitesimal operators (\ref{ben-gen})
rewritten in the canonical form \cite[Sec.~8.4.2.]{Ibr99} and
restricted on the sub-manifold
 \begin{equation}
 t,\ x, \ A^0, \ \ldots , \  A^i, \  \ldots \,.
 \label{eq11}
 \end{equation}
are given by the following expressions
 \begin{equation}
 \begin{array}{c}
 \displaystyle{
 X_1 = \sum\limits_{i=0}^{\infty}\left(
 A^{i+1}_{x}+iA^{i-1}A_{x}^{0}\right)
 \partial_{A^i}\,; \
 X_2 = \sum\limits_{i=0}^{\infty}A^{i}_{x}\partial_{A^i}\,; \
 X_3 = \sum\limits_{i=0}^{\infty}\left( iA^{i-1}-tA^{i}_{x}\right)
 \partial_{A^i}\,;} \\
 \mbox{} \\
 \displaystyle{
 X_4 = \sum\limits_{i=0}^{\infty}\left[
 (i+2)A^{i}-t(A^{i+1}_{x}+iA^{i-1}A^{0}_{x})\right]
 \partial_{A^i}\,; \
 X_5 = \sum\limits_{i=0}^{\infty}\left[(i+2)A^{i}-xA^{i}_{x}\right]
 \partial_{A^i}\,.}
 \end{array}
 \label{gen-moment} \end{equation}
It can be easily checked that infinitesimal operators
(\ref{gen-moment}) are admitted by Benney equations (\ref{ben2})
and it goes without saying that they directly result from the
group analysis of Benney equations (\ref{ben2}). Just in this way
(i.e., using the method of moments) infinitesimal operators
(\ref{ben-gen}) were first obtained in \cite{Kras89} by using
non-canonical form of infinitesimal operators (\ref{gen-moment})
with the subsequent passage to the representation (\ref{ben-gen})
in the space of variables (\ref{var3}).

\subsection{Incompleteness of the point group: statement of the
problem}

It is evident, however, that the subgroup (\ref{gen-moment}) does
not exhaust the complete group symmetry of Benney equations
(\ref{ben2}). The incompleteness of the result (\ref{gen-moment})
is obvious form many points of view. Here we shall only point on
the nonconformity of finite dimension of the algebra
(\ref{gen-moment}) to the infinite set of conservation laws for
Benney equations, and on the infinite extension of the point
symmetry group for Benney equations in the form of (\ref{ben3}),
(\ref{ben4}) with Lagrangian velocity (see also in the context
Chapter 16 in \cite{CRC} where this extension was outlined for
Vlasov kinetic equation in plasma theory). Here of principle
significance for us is the following statement: {\it the group}
(\ref{gen-moment}) {\it is incomplete not only from the standpoint
of Lie-B\"acklund symmetry for Benney equations but also from the
standpoint of the Lie point symmetry}. The validity of the
statement can be proved by direct solving of determining equations
for the first order Lie-B\"acklund group (contact group, that is
not reduced to point one)
\par
 \begin{equation} D_{t}(\mbox{\ae}^i) + D_{x}(\mbox{\ae}^{i+1})
 +iA^{i-1} D_{x}(\mbox{\ae}^0)
 +iA^{0}_x \mbox{\ae}^{i-1} = 0\,, \quad i \geq 0 \,,
 \label{eq13}
 \end{equation}
where coordinates $\mbox{\ae}^i$ of canonical operator
 \begin{equation}
 X=\sum\limits_{i=0}^{\infty} \mbox{\ae}^{i} \partial_{A^i}\,,
 \label{eq14} \end{equation}
depend upon the countered set of group variables
 \begin{equation}
 t, \  x; \  A^0, \ \ldots , \ A^j, \ \ldots \,; \
 A^0_x, \ \ldots , \ A^j_x , \ldots \,; \quad j\geq 0 \,.
 \label{eq15}
 \end{equation}
To prove the above statement one can consider only partial
solutions of determining equations (\ref{eq13})
 \begin{equation}
 \mbox{\ae}^{i}=\eta^i (A^0, \ldots , A^j,\ldots ) \,; \quad i,j\geq 0,
 \label{eq16}
 \end{equation}
that depend upon moments $A^j,\ j\geq 0$, and does not depend upon
$t, \ x$.  It appears that thanks to these infinitesimal operators
(\ref{eq14}), (\ref{eq16}) an infinite extension of the group
(\ref{gen-moment}) takes place. Now the problem is to find these
operators.

\section{Determining equations and their solution}
\nopagebreak[4]
\subsection{General form of the determining equations}
Before proceeding further we write determining equations of
first-order Lie-B\"acklund group, admitted by a more
general (as compared to (\ref{ben2})) infinite system of coupling
equations for functions $A^{i}(t,x)$ with the arbitrary element
$\varphi(A^0)$
 \begin{equation}
  A_{t}^{i} + A_{x}^{i+1} + iA^{i-1}
 \left[ \varphi(A^{0})\right]_x = 0, \qquad i \geq 0 \,.
 \label{ben2a}
 \end{equation}
For the coordinates $\mbox{\ae}^i$ of canonical infinitesimal
operator (\ref{eq14}) the following chains of determining
equations are valid which result from splitting (\ref{ben2a}) with
respect to second derivatives:
 \begin{equation}
 \begin{array}{l}
 \displaystyle{
 {\mbox{\ae}}^{i+1}_{A^0_x} +i\varphi_1
 A^{i-1}{\mbox{\ae}}^{0}_{A^0_x}
 =\sum\limits_{j=0}^{\infty}j\varphi_1 A^{j-1}
 {\mbox{\ae}}^{i}_{A^j_x}\,, \quad i\geq 0\,;}\\
 \displaystyle{
 {\mbox{\ae}}^{i+1}_{A^j_x} +i\varphi_1
 A^{i-1}{\mbox{\ae}}^{0}_{A^j_x} = {\mbox{\ae}}^{i}_{A^{j-1}_x}\,;
 \quad i\geq 0\,, \ j\geq 1\,,}\\
 \mbox{}\\
 \displaystyle{
 {\mbox{\ae}}^{i}_{t} +{\mbox{\ae}}^{i+1}_{x} +i\varphi_1
 A^{i-1}{\mbox{\ae}}^{0}_{x} +A_{x}^{0}\left( i \varphi_1
 {\mbox{\ae}}^{i-1}+ i \varphi_2 A^{i-1} {\mbox{\ae}}^{0} \right)} \\
 \mbox{} \\
 \displaystyle{ \hphantom{{\mbox{\ae}}^{i}_{t} }
 +\sum\limits_{j=0}^{\infty} \left[ i\varphi_1 A^{i-1} A^{j}_{x}
 {\mbox{\ae}}^{0}_{A^j} -\left( A^{j+1}_{x} +j\varphi_1 A^{0}_{x}
 A^{j-1} \right) {\mbox{\ae}}^{i}_{A^j} +A^{j}_{x}
 {\mbox{\ae}}^{i+1}_{A^j} \right]}\\
 \displaystyle{
 \hphantom{{\mbox{\ae}}^{i}_{t} } -\sum\limits_{j=0}^{\infty} j
 A^{0}_{x} \left(\varphi_1 A^{j-1}_{x} +\varphi_2 A^{0}_{x} A^{j-1}
 \right) {\mbox{\ae}}^{i}_{A^j_x} = 0\,, \quad i\geq 0\,.}
 \end{array}
 \label{eqA.1}
 \end{equation}
Here $\varphi_1$ and $\varphi_2$ are the first and the second
derivatives of the function $\varphi$ with respect to its
argument. From the various standpoints at list three distinct
values of the function $\varphi$ are specified. In case $\varphi
(A^0) = A^0 $ we come to kinetic Benney equations (\ref{ben1}),
whereas for $\varphi=a(A^0)^2$ extension of the admitted point
group takes place thanks to projective transformations in
$t,x$-plane (see \cite{Kras89}). For $\varphi=a \ln A^0$ the
corresponding kinetic equation
 \begin{equation}
 f_t+vf_x-a\frac{A^{0}_{x}}{A^0}f_v =0 \,, \quad A^0 =
 \int\limits_{-\infty}^{+\infty} dv\, f\,,
 \label{eqA.2}
 \end{equation}
that gives rise to the discussed system of equations for moments,
is of special interest in plasma theory. It appears as the
equation for the distribution function of plasma ions, while
electrons obey the Boltzmann distribution. More complicated
dependencies of $\varphi(A^0)$ upon $A^0$ can also be of interest
in plasma physics for non-Boltzmann distribution functions for hot
electrons. The equation (\ref{eqA.2}) was studied in details
in~\cite{Gurevich80}.
\par
For the Benney equations (\ref{ben2}) the determining equations
(\ref{eqA.1}) are rewritten in the following form
 \begin{equation}
 \begin{array}{l}
 \displaystyle{ {\mbox{\ae}}^{i+1}_{A^0_x} +i
 A^{i-1}{\mbox{\ae}}^{0}_{A^0_x} -\sum\limits_{j=0}^{\infty}j
 A^{j-1} {\mbox{\ae}}^{i}_{A^j_x}=0\,, \quad i\geq 0\,;}\\
 \displaystyle{ {\mbox{\ae}}^{i+1}_{A^{j+1}_x}
 -{\mbox{\ae}}^{i}_{A^j_x} +iA^{i-1}
 {\mbox{\ae}}^{0}_{A^{j+1}_x}=0\,, \quad i\geq 0\,, \ j\geq 0\,.}\\
 \displaystyle{ {\mbox{\ae}}^{i}_{t}
 +{\mbox{\ae}}^{i+1}_{x} + i A^{i-1}{\mbox{\ae}}^{0}_{x}
 +A_{x}^{0}\left( i {\mbox{\ae}}^{i-1} -\sum\limits_{j=0}^{\infty}
 j A^{j-1} {\mbox{\ae}}^{i}_{A^j} -\sum\limits_{j=0}^{\infty} (j+1)
 A^{j}_{x} {\mbox{\ae}}^{i}_{A^{j+1}_x} \right) }\\
 \displaystyle{ \hphantom{{\mbox{\ae}}^{i}_{t} } + i A^{i-1}
 \sum\limits_{j=0}^{\infty} A^{j}_{x} {\mbox{\ae}}^{0}_{A^{j}}
 +\sum\limits_{j=0}^{\infty} A^{j}_{x} {\mbox{\ae}}^{i+1}_{A^{j}}
 -\sum\limits_{j=0}^{\infty} A^{j+1}_{x} {\mbox{\ae}}^{i}_{A^{j}}
 =0 \,, \quad i\geq 0\,.}
 \end{array}
 \label{eqA.3} \end{equation}

\subsection{Solution of the determining equations}

Under conditions (\ref{eq16}) the determining equations
(\ref{eqA.3}) are splitted and reduced to two infinite chains of
equalities, namely one-dimensional (vector) and two-dimensional
(tensor):
 \begin{equation}
 \begin{array}{l}
 \displaystyle{
 \eta^{i+1}_{A^0} -\sum\limits_{j=0}^{\infty}j
 A^{j-1}\eta_{A^j}^{i} + iA^{i-1}\eta_{A^0}^{0} +i\eta^{i-1} = 0\,,
 \quad i \geq 0\,; } \\ \mbox{} \\
 \displaystyle{
 \eta^{i+1}_{A^{k+1}} -\eta^{i}_{A^{k}} + iA^{i-1}\eta_{A^{k+1}}^{0} = 0\,,
 \quad i \geq 0\,, \ k \geq 0\,.}
 \end{array}
 \label{eq17}
 \end{equation}
The apparent difficulty in analytical solving of the given system
of determining equations (\ref{eq17}) is due to a ``nonlocal"
nature of the second term in the vector chain in the form of an
infinite sum with respect to index $j\geq 0$.  The measure of this
non-locality is characterized by a number of nonzero components of
tensor $\eta^{i}_{j}$.  But in fact in case of an overdetermined
system (\ref{eq17}) we obtain a finite upper value of the
summation index $j < \infty $, which depends upon the other index
$i$ of this tensor. In order to be sure that it is true we shall
first consider the system of determining equations (\ref{eq17}) in
two particular cases, namely for $i =0$ and $i = 1$. In case $i
=0$ we have two coupled determining equations just for two
coordinates $\eta^0$ and $\eta^1$ of the desired infinitesimal
operator (\ref{eq14}), (\ref{eq16}):
\begin{equation}
\eta^{1}_{A^0} -\sum\limits_{j=0}^{\infty}j A^{j-1}\eta_{A^j}^{0}
= 0\,, \qquad \eta^{1}_{A^{k+1}}=\eta^{0}_{A^{k}}\,, \quad  k \geq
0\,. \label{eq18} \end{equation} The second determining equation
in (\ref{eq18}) enables to eliminate the coordinate $\eta^0$ from
the first determining equation and obtain as a consequence of the
system (\ref{eq18}) the following isolated scalar determining
equation for the coordinate $\eta^{1}$ only
\begin{equation}
\eta^{1}_{A^0} -\sum\limits_{j=0}^{\infty}j
A^{j-1}\eta_{A^{j+1}}^{1} = 0\,. \label{eq19} \end{equation} For
$i = 1$ the system (\ref{eq17}) yields coupled equations for three
coordinates $\eta^0$, $\eta^1$ and $\eta^2$
\begin{equation}
\eta^{2}_{A^0} =\sum\limits_{j=0}^{\infty}j A^{j-1}\eta_{A^j}^{1}
- \left(\eta^{0} + A^{0}\eta_{A^0}^{0} \right) , \quad
\eta^{2}_{A^{k+1}} = \eta^{1}_{A^{k}} - A^{0}\eta_{A^{k+1}}^{0} ,
\quad  k \geq 0\,. \label{eq20} \end{equation} Compatibility
conditions for determining equations (\ref{eq20})
\begin{equation}
\eta^{2}_{A^0 A^{k+1}} = \eta^{2}_{A^{k+1} A^0} , \quad k\geq 0\,,
\label{eq21} \end{equation} enables to eliminate the coordinate
$\eta^2$ from (\ref{eq20}) and obtain one more closed vector
determining equation for $\eta^1$ (more precisely, the determining
equation that contains the first and the second derivatives of
$\eta^1$ with respect to $A^i$):
\begin{equation}
\eta^{1}_{A^k A^0} -\sum\limits_{j=0}^{\infty}j
A^{j-1}\eta_{A^{j}A^{k+1}}^{1} - (k+2)\eta^{1}_{A^{k+2}} = 0\,,
\quad k \geq 0\,. \label{eq22} \end{equation} Differentiating the
scalar equality (\ref{eq19}) by moments $A^k$ yields one more
vector corollary for the coordinate $\eta^1$:
\begin{equation}
\eta^{1}_{A^0 A^k} -\sum\limits_{j=0}^{\infty}j
A^{j-1}\eta_{A^{j+1}A^{k}}^{1} - (k+1)\eta^{1}_{A^{k+2}} = 0\,,
\quad k \geq 0\,. \label{eq23} \end{equation} From the
compatibility conditions for two determining equations
(\ref{eq22}) and (\ref{eq23}) it follows that the coordinate
$\eta^1$ does not depend upon the moments $A^i$  which are higher
than $A^1$
\begin{equation} \eta^{1}_{A^{i+2}} = 0\,, \quad i \geq 0\,.
\label{eq24} \end{equation} This formula arises as a result of
mutual subtraction of determining equations (\ref{eq22}) and
(\ref{eq23}) in view of the equality of terms containing summation
over index $ j \geq 0$, that is the corollary of the second
determining equation in (\ref{eq18})
\begin{equation}
\eta^{1}_{A^{j+1} A^{k}} = \eta^{1}_{A^j A^{k+1}}\,, \quad j, \ k
\geq 0\,. \label{eq25} \end{equation} The symmetry of the second
derivatives of $\eta^1$ given by (\ref{eq25}) results from the
tensor form of a compatibility condition for two vector
determining equations from (\ref{eq18}), which differ only in
``sounding" index
\begin{equation}
\eta^{0}_{A^{k}} = \eta^{1}_{A^{k+1}}\,,  \quad \eta^{0}_{A^{j}} =
\eta^{1}_{A^{j+1}}\,, \quad j,\ k \geq 0\,. \label{eq26}
\end{equation} The result (\ref{eq24}) is a milestone on the way
of solving the system of determining equations (\ref{eq17}).
Indeed, in view of (\ref{eq24}) the second determining equation of
the system (\ref{eq18}) yields the requirement of independence of
the coordinate $\eta^0$ upon all higher moments, that differ from
$A^0$
\begin{equation}
\eta^{0}_{A^{i+1}} = 0\,, \quad i \geq 0\,, \label{eq27}
\end{equation} and with (\ref{eq27}) in mind the tensor chain from
(\ref{eq17}) is simplified in such a way
\begin{equation}
\eta^{i+1}_{A^{k+1}} = \eta^{i}_{A^{k}}\,,  \quad i\geq 0 ,\ k
\geq 0\,, \label{eq28} \end{equation} that enables to modify the
equality (\ref{eq24}) in the sense that any coordinate $\eta^i$ of
the desired operator (\ref{eq14}), (\ref{eq16}) does not depend
upon any moments $A^j$, higher than $A^i$
\begin{equation}
\eta^{i}_{A^{i+k+1}} = 0\,, \quad i \geq 0\,, \ k\geq 0\,.
\label{eq29} \end{equation} The last equality sets a finite upper
limit $i\geq j$ to the summation index $j\geq 0$ in the second
(nonlocal) term of the left hand side of the vector determining
equation of the system (\ref{eq17})
\begin{equation}
\eta^{i+1}_{A^0} -\sum\limits_{j=0}^{i}j A^{j-1}\eta_{A^j}^{i} +
iA^{i-1}\eta_{A^0}^{0} +i\eta^{i-1} = 0\,, \quad i \geq 0\,.
\label{eq30} \end{equation} The use of one of the equalities
(\ref{eq28})
\begin{equation}
\eta^{0}_{A^{0}} = \eta^{i}_{A^{i}}\,,  \quad i \geq 0
\label{eq31} \end{equation} enables to rewrite the chain of
determining equations (\ref{eq30}) in even a more simple way
\begin{equation}
\eta^{i+1}_{A^0} -\sum\limits_{j=0}^{i-1}j A^{j-1}\eta_{A^j}^{i}
+i\eta^{i-1} = 0\,, \quad i \geq 0\,, \label{eq32} \end{equation}
whereas one of the corollaries of determining equations
(\ref{eq28}), that results for $i=k+1$ in combination with
(\ref{eq27})
\begin{equation}
\eta^{i+1}_{A^{i}} = 0\,, \quad i \geq 0\,, \label{eq33}
\end{equation} lowers the upper value of the summation index
$j\geq 0$ in (\ref{eq32}) by unit
\begin{equation}
\eta^{i+1}_{A^0} -\sum\limits_{j=0}^{i-2}j A^{j-1}\eta_{A^j}^{i}
+i\eta^{i-1} = 0\,, \quad i \geq 0\,. \label{eq34} \end{equation}
Collecting the arising determining equations (\ref{eq27}),
(\ref{eq28}) and (\ref{eq33}), (\ref{eq34}) we arrive to a much
more simplified (but equivalent) formulation of the system
(\ref{eq17}), which really is integrated below
\begin{equation}
\begin{array}{l}
\displaystyle{ \eta^{i+1}_{A^0} -\sum\limits_{j=0}^{i-2}j
A^{j-1}\eta_{A^j}^{i} +i\eta^{i-1} = 0\,, \quad \eta^{i+1}_{A^{i}}
= 0\,, \quad i \geq 0\,; } \\ \mbox{}\\ \displaystyle{
\eta^{i+1}_{A^{k+1}} = \eta^{i}_{A^{k}}\,, \quad
\eta^{i}_{A^{i+k}} = 0\,, \quad i \geq 0\,,\ k \geq 0\,. }
\end{array}
\label{eq35} \end{equation} The last of the four equalities in
(\ref{eq35}) is strengthened in comparison with (\ref{eq29})
thanks to the condition
\begin{equation}
\eta^{i}_{A^{i}} = 0\,, \quad i \geq 0\,, \label{eq36}
\end{equation} which can be obtained as follows. For example, the
first two equalities (\ref{eq36}) for $i=0$ and $i=1$ result as a
corollary of (\ref{eq31}) and compatibility conditions
\begin{equation}
\eta^{3}_{A^{0} A^{1}} = \eta^{3}_{A^1 A^{0}} \label{eq37}
\end{equation} for the first derivatives of the coordinate
$\eta^3$ with respect to moments $A^0$ and $A^1$ (see (\ref{eq34})
for $i=1$ and $i=2$)
\begin{equation}
\eta^{3}_{A^{0}}  = -2 \eta^{1} \,, \quad \eta^{3}_{A^{1}}  =
\eta^{2}_{A^0}  = - \eta^{0} \,. \label{eq38} \end{equation} After
that, the validity of the remaining equalities (\ref{eq36})
becomes obvious thanks to~(\ref{eq31}).
\par
Before proceeding to enumerating all solutions of the system of
determining equations (\ref{eq35}), we present here yet another
form of the chain (\ref{eq34})
 \begin{equation}
 \eta^{i+1}_{A^0} -\sum\limits_{j=0}^{i-2}j A^{j-1}\eta_{A^0}^{i-j}
 +i\eta^{i-1} = 0\,, \quad i \geq 0\,.
 \label{eq39}
 \end{equation}
This form can be employed to clarify the general structure of
these solutions on basis of the corresponding generating
functions.

\subsection{Discussion of the solution of the determining equations}
\par
The integrability procedure in itself for determining equations
(\ref{eq35}) is of no difficulties. For example the first six
coordinates $\eta^i$ $(0 \leq i \leq 5)$ of the desired
infinitesimal operator (\ref{eq14}), (\ref{eq16}) are given by the
following formulas for the general solutions of determining
equations (\ref{eq35}) that depend upon six arbitrary constants
$C^j$ $(0 \leq j \leq 5)$ and are described by polynomials in
moments $A^l$
\begin{equation}
\begin{array}{l}
\displaystyle{ \eta^0=C^0, \quad \eta^1=C^1, \quad \eta^2=C^2-C^0
A^0,  \quad \eta^3=C^3-2C^1 A^0 -C^0 A^1, } \\ \mbox{}\\
\displaystyle{ \eta^4=C^4-3C^2 A^0 - 2 C^1 A^1 +C^0
\left[-A^2+(A^0)^2 \right]\,, }\\ \mbox{}\\ \displaystyle{ \eta^5
= C^5 - 4C^3 A^0 - 3 C^2 A^1 + C^1 \left[-2A^2+3(A^0)^2 \right] }
\displaystyle{ + C^0 \left(- A^3 + 2A^0 A^1\right)\,.}
\end{array}
\label{eq40} \end{equation} It appears that the polynomial
dependence of any solution $\eta^i$ of determining equations
(\ref{eq35}) upon moments $A^j$ is a general property of
components of the vector $\eta^i$ for any $i\geq 0$. The example
(\ref{eq40}) demonstrates that the procedure of obtaining
solutions of determining equations (\ref{eq35}) is reduced to
their enumeration. To be concrete, we assume the following scheme
of indicating of the $k$-th basic solution $\eta^{i}_{k}$ of
determining equations (\ref{eq35}) for the coordinate $\eta^i$:
 \begin{equation}
 \eta^{i}_{k}= \left\{
 \begin{array}{ll}
 0, & i < k; \\ 1, & i = k; \\ 0, & i = k+1; \\
 \end{array}
 \right. \quad \left[ \eta^{i}_{k} \right] = i-k, \ i \geq k+2; \quad
 i,\,k \geq 0\,.
 \label{eq41}
 \end{equation}
In the solutions (\ref{eq40}) this scheme demands quit definite
choice of values of integration constants $C^j$ in the form of
Kronecker symbols
\begin{equation}
C^j=\delta_{j k}\,; \quad j,\, k \geq 0\,. \label{eq42}
\end{equation} The last of the four equalities for $\eta^i_{k}$ in
(\ref{eq41}) (in square brackets) indicates the homogeneity degree
$(i-k)$ of the polynomial ``tail" of the solution $\eta^i$ for
$i\geq k+2$ in accordance with the attributed to any of the
moments $A^i$ of the order $i$ the homogeneity degree, which is
equal to positive number $(i+2)$ (see e.g. \cite{Kup77})
\begin{equation} \left[ A^{i} \right] = i+2 \,, \quad  i \geq 0\,.
\label{eq43} \end{equation} For instance, the component $\eta^5_1$
of the basis solution $\eta^i_1$ of determining equations
(\ref{eq35}) in accordance with (\ref{eq40}), (\ref{eq41}) and
(\ref{eq43}) has the homogeneity degree equal to four
\begin{equation} \eta^5_1 = - 2 A^2  + 3(A^0)^2 \,; \quad \left[
\eta^{5}_{1} \right] = 4\,. \label{eq44} \end{equation} The
indexing of the presented infinite (countable) vectors $\eta^i$ by
one more integral number $k\geq 0$ yields the desired
representation of all linear independent solutions of determining
equations (\ref{eq35}) in the form of tensor of the second rank
(matrix) $\eta^i_k$, in which the lower index $k\geq 0$ indicates
the index of the basis infinitesimal operator in the general
element of an infinite Lie algebra under consideration
\begin{equation}
X=\sum\limits_{i,k=0}^{\infty} C^k \, \eta^{i}_{k} \,
\partial_{A^i}\,, \label{eq45} \end{equation} Under the conditions
(\ref{eq41}) the integration of determining equations (\ref{eq35})
for the given basis vector $\eta_{k}^{i}$ for a fixed value $k\geq
0$ is carried out with boundary conditions, that are imposed by
requirements (\ref{eq41}) in a single way.
\par
The representation of matrix $\eta^{i}_{k}$ for different lines
are as follows ($i$ is the column number, $k$ is the line number)
\begin{equation}
\eta^{i}_{k}=\{ 0, \ldots\,, 0,\, 1,\,0,\, -(k+1)A^0,\, -(k+1)A^1,
\ldots \} \,. \label{eq46} \end{equation} Here zeroes preceding
unity describe matrix elements, which exist only for $i<k$, i.e.
which are located below the principle diagonal $i=k$, that
contains only units. The first nearest upper off-diagonal $i=k+1$
also contains only zeroes. Expressions for elements from the
second $i=k+2$ and the third $i=k+3$ upper off-diagonals are given
in (\ref{eq46}) explicitly: they contain monomials, the
homogeneity degree of which is equal to $2$ and $3$ respectively,
while the numerical coefficient $(k+1)$ is defined by the line
number.
\par
In general, any one of the nonzero off-diagonals $i=k+s$ with the
number $s\geq 2$ is presented by polynomials with the homogeneity
degree equal to $s$. This ``line scheme" (\ref{eq46}) is readily
illustrated by a pictorial rendition of elements of the high left
block of the discussed matrix ($0\leq i \leq 5$, $0 \leq k \leq
3$)
 \begin{equation}
 \eta^{i}_{k}=\left(
 \begin{array}{c c c c c c c}
 1      & 0   & -A^0 & -A^1 & -A^2 + (A^0)^2 & -A^3+2 A^0 A^1  &
 \ldots \\ 0      & 1   &   0  &-2A^0 &      -2A^1     & -2A^2+3
 (A^0)^2  &  \ldots \\ 0      & 0   &   1  &  0   &      -3A^0
 &     -3A^1        &  \ldots \\ 0      & 0   &   0  &  1   &
 0      &      -4 A^0      &  \ldots \\ \ldots & \ldots  & \ldots
 & \ldots  & \ldots & \ldots &  \ldots
 \end{array} \right)
 \label{eq47} \end{equation}
As a more illustrative example we present here the element
$\eta^i_1$ of the matrix (\ref{eq46}) with sufficiently high
column number $i=10$ and the homogeneity degree $9$, that is
located in the line with $k=1$ (the second from above)
 \begin{equation}
 \begin{array}{ll}
  \eta^{10}_{1}= & -2 A^7 +6 A^5 A^0 +6 A^4 A^1 +6 A^3 A^2
   -12 A^3 (A^0)^2 \\[2ex]
      & -24 A^2 A^1 A^0 -4 (A^1)^3 +20 A^1 (A^0 )^3\,.
 \end{array}
 \label{eq48} \end{equation}

\subsection{Illustrative example for matrix elements}
A much more comprehensive idea of definite  expressions of matrix
elements $\eta^i_k$ is given by the following list of elements
(with the previous result included) of the first $11$ columns
($0\leq i \leq 10$) and $4$ lines ($0\leq k \leq 3$) of matrix
$\eta^i_k$, which define the $k$-th basic solution of determining
equations (\ref{eq35}) for vectors $\eta^i_k$ of the canonical
infinitesimal operator (\ref{eq14}), (\ref{eq16}). The lower index
``$k$" is omitted for simplicity.
\smallskip
\par
\noindent
\begin{equation}
\begin{array}{ll}
0) & k=0; \ \eta^0=1, \ \eta^1=0, \ [\eta^i] =i, \ i \geq 2.
   \mbox{\hphantom{x} \hfill} \\
\mbox{}  & \eta^2=-A^0\,, \\ \mbox{}  & \eta^3=-A^1\,, \\ \mbox{}
& \eta^4=-A^2 + (A^0)^2\,, \\ \mbox{}  & \eta^5=-A^3 + 2 A^0 A^1
\,, \\ \mbox{}  & \eta^6=-A^4 + 2 A^0 A^2 + (A^1)^2 - (A^0)^3 \,,
\\ \mbox{}  &  \eta^7=-A^5 + 2 A^0 A^3 + 2 A^2 A^1 - 3 A^1 (A^0)^2
\,, \\ \mbox{}  &  \eta^8=-A^6 + 2 A^0 A^4 + 2 A^3 A^1 + (A^2)^2 -
3 A^2 (A^0)^2\\ \mbox{}  &  \hphantom{\eta^9=} - 3 A^0 (A^1)^2
+(A^0)^4 \,, \\ \mbox{}  &  \eta^9=-A^7 + 2 A^0 A^5 + 2 A^4 A^1 +
2 A^3 A^2 - 3 A^3 (A^0)^2 \\ \mbox{}  &  \hphantom{\eta^9=} - 6
A^0 A^1 A^2 - (A^1)^3 + 4 A^1 (A^0)^3 \,, \\ \mbox{}  &
\eta^{10}=-A^8 + 2 A^0 A^6 + 2 A^5 A^1 + A^4 [2 A^2 - 3 (A^0)^2 ]
             +A^3[A^3 - 6 A^0 A^1] \\
\mbox{}  &  \hphantom{\eta^10=}
       + A^2 [-3 (A^1)^2 -3 A^0 A^2 + 4 (A^0)^3]+6(A^1)^2(A^0)^2 - (A^0)^5 \,.
\end{array}
\label{eqA.8} \end{equation}
\smallskip
\par
\noindent
\begin{equation}
\begin{array}{ll}
 1) & k=1; \ \eta^0=0, \ \eta^1=1, \ \eta^2=0, \ [\eta^i ]=i-1, \ i
 \geq 3 .
   \hbox to 3cm {\mbox{} \hfill} \\
 \mbox{}  &\eta^3=-2A^0\,, \\ \mbox{}  &\eta^4=-2A^1\,, \\ \mbox{}
 &\eta^5=-2A^2 + 3(A^0)^2\,, \\ \mbox{}  &\eta^6=-2A^3 + 6 A^0 A^1
 \,, \\ \mbox{}  &\eta^7=-2A^4 + 6 A^0 A^2 + 3(A^1)^2 - 4(A^0)^3
 \,, \\ \mbox{}  &\eta^8=-2A^5 + 6 A^0 A^3 + 6 A^2 A^1 - 12 A^1
 (A^0)^2 \,, \\ \mbox{}  &\eta^9=-2A^6 + 6 A^0 A^4 + 6 A^3 A^1 +
 A^2[3A^2 - 12 (A^0)^2] \\ \mbox{}  &\hphantom{\eta^9=}
           - 12 A^0 (A^1)^2 + 5(A^0)^4 \,, \\
 \mbox{}  &\eta^{10}=-2A^7 + 6 A^0 A^5 + 6 A^4 A^1 + 6 A^3 [A^2 - 2
 (A^0)^2] \\ \mbox{}  &\hphantom{\eta^10=}
          - 24 A^0 A^1 A^2 + A^1[-4(A^1)^2 + 20 (A^0)^3] \,.
 \end{array}
 \hphantom{\textstyle{\hfill}} \label{eqA.9} \end{equation}
 \smallskip
 \par
 \noindent
 \begin{equation}
 \begin{array}{ll}
 2) & k=2; \ \eta^0=0, \ \eta^1=0,  \ \eta^2=1, \eta^3=0, \
 [\eta^i]=i-2,  \ i \geq 4 .
   \hbox to 1.5cm {\mbox{} \hfill} \\
 \mbox{}  &\eta^4=-3A^0\,, \\ \mbox{}  &\eta^5=-3A^1\,, \\ \mbox{}
 &\eta^6=-3A^2 + 6(A^0)^2\,, \\ \mbox{}  &\eta^7=-3A^3 + 12 A^0 A^1
 \,, \\ \mbox{}  &\eta^8=-3A^4 + 12 A^0 A^2 + 6(A^1)^2 - 10(A^0)^3
 \,, \\ \mbox{}  &\eta^9=-3A^5 + 12 A^0 A^3 + 12 A^2 A^1 - 30 A^1
 (A^0)^2 \,, \\ \mbox{}  &\eta^{10}=-3A^6 + 12 A^0 A^4 + 12 A^3 A^1
 + 6(A^2)^2 \\ \mbox{}  &\hphantom{\eta^10=}
           - 30 A^0 (A^1)^2 +15(A^0)^4 -30 A^2 (A^0)^2\,.
 \end{array}
 \label{eqA.10}
 \end{equation}
 \smallskip
 \par
 \noindent
 \begin{equation}
 \begin{array}{ll}
 3) & k=3; \ \eta^0=0, \ \eta^1=0, \ \eta^2=0, \ \eta^3=1, \
 \eta^4=0,
    \ [\eta^i]=i-3,  \ i \geq 5.
 \hbox to 0.1cm {\mbox{} \hfill} \\ \mbox{}  &\eta^5=-4A^0\,, \\
 \mbox{}  &\eta^6=-4A^1\,, \\ \mbox{}  &\eta^7=-4A^2 + 10(A^0)^2\,,
 \\ \mbox{}  &\eta^8=-4A^3 + 20 A^0 A^1 \,, \\ \mbox{}
 &\eta^9=-4A^4 + 20 A^0 A^2 + 10(A^1)^2 - 20(A^0)^3 \,, \\ \mbox{}
 &\eta^{10}=-4A^5 + 20 A^0 A^3 + 20 A^2 A^1 - 60 A^1 (A^0)^2 \,.
 \end{array}
 \label{eqA.11} \end{equation}

\section{Conclusion}

This paper presents a result of calculation of the infinite
(countable) part of Lie point group admitted by the system of
Benney equations -- moment equations (\ref{ben2}). In standard
(non-canonical representation) the point Lie group of Benney
equations (\ref{ben2}) is described by the infinitesimal operator
 \begin{equation} X=\xi^1 \partial_t
    + \xi^2 \partial_x
    +\sum\limits_{i=0}^{\infty} \eta^i \frac{\partial}{\partial {A^i}}\,.
 \label{eq49} \end{equation}
where coordinates $\xi$ and $\eta$ obey the system of determining
equations
 \begin{equation}
 \begin{array}{l}
 \displaystyle{ \eta^{i+1}_{A^0}
              -\sum\limits_{j=0}^{\infty}j A^{j-1}
              {\eta}^{i}_{A^j_x}
              +i\eta^{i-1}+iA^{i-1}\left( \eta^0_{A^0}
              +\xi^1_t-\xi^2_x \right) }
 \displaystyle{
             +(i+1)A^i\xi^1_x - \xi^2_t\delta_{i,0} = 0 , }
 \\ \mbox{}\\
 \displaystyle{ {\eta}^{i+1}_{A^{k+1}}
              -{\eta}^{i}_{A^{k}} +iA^{i-1} \left(
               {\eta}^{0}_{A^{k+1}}
              + \xi^1_x\delta_{0,k} \right)}
 \displaystyle{
             +(\xi^1_t - \xi^2_x )\delta_{i,k}
             +\xi^1_x \delta_{i+1,k}
             - \xi^2_t\delta_{i,k+1} =0, }
 \\
 \mbox{}\\ \displaystyle{
            {\eta}^{i}_{t}
           +{\eta}^{i+1}_{x}
           +i A^{i-1}{\eta}^{0}_{x} = 0\,, \qquad i, \ k \geq 0\,.}
 \end{array}
 \label{eq50} \end{equation}
Determining equations (\ref{eq50})
result from (\ref{eqA.3}) in account of relationships between
coordinates of infinitesimal operators (\ref{eq49}) and
(\ref{eq14})
 \begin{equation}
 {\mbox{\ae}}^i = \eta^i +\xi^1 (A^{i+1}_{x} +i A^{i-1}A_x^0)
 -\xi^2 A^i_x \,.
 \label{eq51} \end{equation}
Infinitesimal operators (\ref{gen-moment}), that were presented
above, gives rise to the following coordinates
 \begin{equation}
 \begin{array}{c} \xi^1=K^4+K^5t\,, \quad \xi^2=K^1+K^2t + K^3 x \,,  \\
 \mbox{}\\ \eta^i =iA^{i-1} K^2 + (i+2)A^i (K^3-K^5)\,.
 \end{array}
 \label{eq52} \end{equation}
The problem of finding coordinates of the operator (\ref{eq49})
was first treated in \cite{Kras89}, where only these solutions,
namely (\ref{ben-gen}), (\ref{gen-moment}) and (\ref{eq52}), were
described.  The main result of our paper is that point symmetries
of Benney equations (\ref{ben2}) are exhausted by formulas
(\ref{gen-moment}) and solutions of determining equations
(\ref{eq35}), i.e.  determining equations (\ref{eq50}) do not have
any other solutions. Solutions of determining equations
(\ref{eq35}) which are responsible for the infinite part of the
point group probably have not been known so far.
\par
As a next step it seems intriguing to generalize the result
(\ref{eq51}), i.e. to find the first order Lie-B\"acklund group
admitted by Benney equations (\ref{ben2}) with coordinates
$\mbox{\ae}^{i}$ of the canonical infinitesimal operator
(\ref{eq14}), that has the linear form
 \begin{equation}
 \mbox{\ae}^{i}=\eta^i +
 \sum\limits_{j=0}^{\infty}\eta^{i,j}A^{j}_{x}\,, \qquad i\geq 0\,.
 \label{eq53} \end{equation}
Though the unique existence of the linear form (\ref{eq53}) as
well as the complete solution of determining
equations{\footnote{For simplicity these equations are omitted
here.}} for the tensor $\eta^{i,j}$ has not yet been obtained, all
known facts are in agreement with this linear form. In particular,
results of \cite{Kup77,Kup78} mentioned above are consistent with
the following expression for the tensor $\eta^{i,j}$ of the linear
form
 \begin{equation}
 \eta^{i,j}_{s} = \sum\limits_{k=0}^{\infty}kH^{s}_{A^k}\delta_{i+k,j+1} +
 s\sum\limits_{k=0}^{s-j-2}(i+k)A^{i+k-1}H^{s-1}_{A^{j+k+1}}\,;
 \quad i,\ j,\ s \geq 0\,.
 \label{eq54} \end{equation}
Here $s$ is the number of the basis solution (similar to that used
for $\eta^i$ in (\ref{eq47})), $H^s$ is a polynomial of the
homogeneity degree $(s+2)$ in moments $A^i$. Compatibility
conditions for determining equations for the tensor $\eta^{i,j}$
give rise to many relationships for $H^s$, for example
 \begin{equation}
 \sum\limits_{j=0}^{\infty}j A^{j-1} H^{s}_{A^j} = sH^{s-1}\,,
 \quad  s \geq 0\,.
 \label{eq55} \end{equation}
An explicit form
for the polynomial $H^7$ is presented below just to illustrate the
aforesaid
 \begin{equation}
 \begin{array}{l}
 H^{7} = A^{7} +7A^{5}A^{0} +7A^{4}A^{1} +7A^{3}A^{2}
 +21A^{3}(A^{0})^2 +42 A^{2} A^{1} A^{0} \\ \hphantom{H^7 = } +7
 (A^{1})^3 +35 A^1(A^{0})^3 \,.
 \end{array}
 \label{eq56} \end{equation}
Comparison between formulas (\ref{eq48}) and (\ref{eq56}) shows
that they differ only in numerical values (and signs) of
coefficients. The generating function for polynomials $H^s$ is
given in \cite{Kup77,Kup78}. So constructing of a recursion
operator, which transforms the linear form (\ref{eq51}) for point
group to the linear form (\ref{eq53}) is of principal interest.
\par
To complete the conclusion we present formulas for the infinite
dimensional algebra of a Lie point group admitted by the system of
equations (\ref{ben5})
 \begin{equation}
 \begin{array}{l}
 \displaystyle{
 X_1 = \xi (t) \partial_{t} + \frac{x}{2}\,\xi_t \partial_{x} +
       \frac{1}{2} \left( x \xi_{tt} - v \xi_{t}\right) \partial_{v} } 
  \displaystyle{
     - \frac{g}{2}\,\xi_{t}\partial_{g}
     - \frac{1}{4}  \left( 2(g+2h)\xi_{t} +
       x^2 \xi_{ttt} \right)\partial_{h} \,,}\\
  \mbox{} \\
 \displaystyle{
 X_2 = \chi (t) \partial_{x} + \chi_t \partial_{v}
        - x \chi_{tt} \partial_{h} \,,} \quad
 \displaystyle{
 X_3 = x \partial_{x} + v\partial_{v} + g\partial_{g}
      + (g+2h)\partial_{h}\,,}\\
 \mbox{} \\
 \displaystyle{
 X_4 = f \partial_{f} + g\left( \partial_{g}
      - \partial_{h} \right) \,,} \quad
 \displaystyle{
 X_5 =  \partial_{f} + v\left( \partial_{g}
        - \partial_{h} \right) \,,} \quad
 \displaystyle{
 X_6 =  \mu(t) \partial_{h} \,,}\\
 \mbox{} \\
 \displaystyle{
 X_7 =  G (t,x,g+h) \left( \partial_{g}
      - \partial_{h} \right) \,.}
 \end{array}
 \label{gen-dif}
 \end{equation}
Here in (\ref{gen-dif}) $\xi(t)$, $\chi(t)$, $\mu(t)$ and
$G(t,x,g+h)$ are arbitrary functions of their arguments. Using the
procedure described in section \ref{prolong} it is easily checked
that prolongation of generators (\ref{ben-gen}) on nonlocal
variables $g$ and $h$ produce generators that directly follow from
(\ref{gen-dif}). Namely,  the prolongation of $X_1$ from
(\ref{ben-gen}) gives $X_1$ provided $\xi = 1$, and $X_2$ from
(\ref{ben-gen}) gives $X_2$ provided $\chi = 1$. Next,
prolongation of $X_3$ from (\ref{ben-gen}) gives $X_2$ provided
$\chi = t$, and $X_4$ from (\ref{ben-gen}) gives $X_1 - X_3/2 -
X_4$ provided $\xi = t$. At last, prolongation of $X_5$ from
(\ref{ben-gen}) gives $X_3 + X_4$.
\par
It should be noted that on one hand the transition from nonlocal
Benney equations (\ref{ben1}) to their ``potential" analog does
not introduce additional independent variables that relate to
restriction onto the boundary sets (compare with
\cite{Chetv-AMST95,Chetv-AAM95}). This fact reflects the property
of the fast decay of the distribution function $f$ in Benney
equations at the infinity, $f(+\infty)= f(-\infty) = 0$. On the
other hand the generators (\ref{gen-dif}) give one more evidence
in favor of constructing symmetry of nonlocal equations using
various representations and different approaches.

\renewcommand{\thesection}{}
\section{Acknowledgements}
 This work is supported in part by a grant from the National Research
 Foundation (NRF) of South Africa and by the Department of Mathematics,
 Blekinge Institute of Technology. One of
the authors (VFK) is also supported from the RFBR grant
99-01-00232, 00-15-96691. We thank our colleagues Sergey Meleshko
and Rafail Gazizov for their valuable comments.

\end{document}